\documentstyle[12pt]{article}

\begin {document}

\begin{flushright}
OITS-558\\
October 1994\\
\end{flushright}

\vskip1cm
\begin{center}
{\large {\bf FACTORIAL MOMENTS}}
\vskip .25cm
{\large {\bf  OF CONTINUOUS ORDER}}
\vskip .75cm
 {\large {\bf  Rudolph C. Hwa}}
\vskip.5cm
 {Institute of
Theoretical Science and Department of Physics\\
University of Oregon,
Eugene, OR 97403}
\end{center}
\vskip1cm
\centerline{\bf Abstract}
\vskip.2cm
\begin{quote}
\quad \quad The normalized factorial moments $F_q$ are continued to
noninteger values of the order $q$, satisfying the condition that the
statistical fluctuations remain filtered out.  That is, for Poisson
distribution
$F_q = 1$ for all $q$.  The continuation procedure is designed with
phenomenology and data analysis in mind.  Examples are given to show how
$F_q$ can be obtained for positive and negative values of $q$.  With $q$
being continuous, multifractal analysis is made possible for multiplicity
distributions that arise from self-similar dynamics.  A step-by-step
procedure of the method is summarized in the conclusion.
\end{quote}

\vskip1cm

\section{Introduction}\label{I}

\indent Continuation of the normalized factorial moments $F_q$  to
arbitrary, noninteger values of $q$ is not just a mathematical problem.
It has high phenomenological significance, and provides a powerful
method to analyze experimental data that can reveal aspects about
multiplicity fluctuations hitherto unexplored.

A historical review of the problem is in order.  Bia\l as and Peschanski
\cite{bp} first introduced $F_q$ as a means of studying scaling behavior of
multiplicity fluctuations as a function of the resolution scale $\delta$, a
subject usually referred to as intermittency.  Although a large part of
the effect seen in the data has recently been found to be due to
Bose-Einstein correlation among like-charge particles \cite{cr,aga},
intermittency at a weaker level is still present for the unlike-charge
particles and its origin remains to be clarified \cite{agakit}, especially for
$q > 2$.  There is far more dynamical information about high-energy
collisions than can be uncovered by studying two-particle correlation
only.  With that point of view forming the basis of our discussion here,
we now outline the problems associated with the use of $F_q$.

Let us first recall the definition of $F_q$:
    \begin{eqnarray}
F_q = {\left<n \left(n - 1\right)\cdots \left(n - q + 1\right) \right>  \over
\left< n\right>^q }
\label{1}
    \end{eqnarray}
where $\left<\cdots\right>$ denotes an average weighted by the multiplicity
distribution $P_n$.  The most outstanding property of $F_q$ discovered in Ref.
\cite{bp} is that it filters out the statistical fluctuations, so any
nontrivial
behavior of $F_q$ is a direct indication of some features about the dynamics of
particle production.  A quick review of that will be given in the beginning of
the next section. Another significant aspect about $F_q$ is that an event can
contribute to (\ref{1}) only if $n \geq q$.  Thus for small $\delta$ where
$\langle n \rangle$ is small in a bin, only rare events with high spikes ($n
\geq q$) contribute.  That is why it is sometimes said that intermittency
measures spiky events.  There are,  however, disadvantages in the use of
$F_q$.  A corollary to the ability to select spiky events is its inability to
extract any dynamical information about dips.  It is by now generally
recognized that rapidity gaps, like voids in galactic structure, are important
to study.  Those are, of course, large dips.  In nuclear collisions where
multiplicity per bin is large, unusual dips, which can be small but deep, are
as significant as unusual spikes.  For such fluctuations it is necessary to
study $F_q$ for $q < 1$, especially negative $q$.  Furthermore, for
multifractal analysis of multiparticle production the continuation of $F_q$ to
noninteger values of $q$ is necessary in order to allow differentiation with
respective to $q$.  These studies cannot be done, if $F_q$ is defined as in
(\ref{1}).

A method to investigate moments of arbitrary order was suggested several
years ago in terms of the $G$ moments \cite{hw}.  It was later modified to
achieve better power-law behavior \cite{hw2}.  However, in overcoming the
defects of $F_q$, the $G$ moments fail to retain the principal attribute of
$F_q$, i.e. the screening of statistical fluctuations.  Subtraction of the
statistical component has to be done by hand \cite{hw3}.  Since that can be
achieved only by simulation, the method is not elegant.  But it has provided
the first glimpses into the multifractal structure of particle production.

In this paper we describe a method that retains both attributes:  it
eliminates statistical fluctuations and is defined for continuous $q$.
Although the mathematical technicalities involved may at first sight appear
to be of theoretical interest only, the method is developed with
phenomenology in mind.  The purpose of the program is to extract
quantitative information about dynamical fluctuations from the
experimental data and to present it in a form suitable for comparison with
theoretical predictions.  Thus the problem of data analysis has not been
overlooked in favor of mathematical expediency in the hope that the
procedure can be readily accessible to direct experimental application.

\section{The Problem}\label{II}

We first review the virtue of factorial moments.  To say that $F_q$ filters out
statistical fluctuation, one first assumes that the latter enters the
multiplicity
distribution as a convolution with the dynamical component
  \begin{eqnarray}
P_n =  S \otimes D
\label{2}
  \end{eqnarray}
where $S$ represents the statistical component, which we take to be
the Poisson distribution ${\cal P}^{(0)}_n$, and $D$ is the dynamical
distribution.  More specifically, (\ref{2}) implies
  \begin{eqnarray}
P_n = \int^{\infty}_0 dt \, {t^n \over n!}\,\, e^{-t}\, D(t)
\quad ,
\label{3}
  \end{eqnarray}
Let the numerator of (\ref{1}) be denoted by $f_q$, i.e.
  \begin{eqnarray}
f_q = \sum^{\infty}_{n=q} \, {n! \over (n - q) !} \, P_n \quad .
\label{4}
  \end{eqnarray}
which is well defined for $q$ being a positive integer.  Substituting (\ref{3})
in (\ref{4}) and performing the summation yield
  \begin{eqnarray}
f_q = \int^{\infty}_0 dt \, t^q \,  D(t) \quad .
\label{5}
  \end{eqnarray}
This is the $q$th moment of the dynamical $D(t)$ and is free of statistical
contamination.  Since $f_1 = \langle n \rangle$, we have
 \begin{eqnarray}
F_q = f_q /f^q_1
 \, .
 \label{6}
     \end{eqnarray}
If $P_n$  is a Poisson distribution, then $D(t) = \delta (t - \langle n
\rangle)$
and $f_q = \langle n \rangle^q$ so
     \begin{eqnarray}
F_q = 1 \, , \quad \mbox{for integer} \quad  q \geq 1 \, .
\label{7}
    \end{eqnarray}
Because of this trivial result, one can state that any nontrivial $F_q$ reveals
the nontrivial properties of $D(t)$.

To generalize $F_q$ to noninteger $q$, it must first be recognized that there
is no unique continuation to complex $q$.  Since $P_n$  must vanish as $n
\rightarrow \infty$ (in fact, it must vanish for $n > N$ for some finite $N$ at
any finite collision energy), there are only a finite number of the $F_q$
moments defined at integer $q$ values.  Without an accumulation of $F_q$ at
infinite $q$, unique continuation to noninteger $q$ is not possible.  Put
differently, we can add to $F_q$ any arbitrary function that vanishes at the
finite range of integer $q$ where $F_q$ is specified and generate another
function $F_q$ at noninteger $q$.

A simple way of continuing (\ref{4}) to arbitrary $q$ is to replace the
factorials by gamma functions, i.e.
\begin{eqnarray}
f_q = \sum^{\infty}_{n=0} \, {\Gamma (n + 1) \over \Gamma (n - q + 1)}
\, P_n
 \, .
\label{8}
  \end{eqnarray}
Procedures similar to (\ref{8}),  such as continuing $d^qG(z)/dz^q$ to
fractional $q$ \cite{old}, have been considered
previously \cite{blaz}-\cite{dr}.  Since $\Gamma (z)$ has poles and oscillates
rapidly among those poles when $z \leq 0$, $F_q$ as defined in (\ref{8})
oscillates between large positive integers of $q$ and is highly suppressed at
large negative $q$.  The question is whether one wants that kind of behavior
at noninteger values of $q$.  If not, what are the guidelines by which one
makes alternative choices of the continuation schemes?

In our view the only guideline is the primary rationale for considering
factorial moments in the first place.  And that is the elimination of
statistical fluctuation at all $q$, not just at integer values of $q$.  If one
substitutes Poisson distribution into (\ref{8}), one will find that $F_q \not=
\langle n \rangle^q$.  Fig. 1 (a) and (b) show the results for $F_q$ when
     \begin{eqnarray}
P_n = {\cal P}^{(0)}_n =  {\left< n\right>^n \over n !} e ^{-\left< n\right>}
\quad ,
\label{9}
  \end{eqnarray}
for $\langle n \rangle = 6$.  Although (\ref{7}) remains true for
positive integers of $q$, $F_q$ is by no means equal to one for all
$q$.  The oscillations have larger amplitudes at high $q$, as revealed in
Fig.\ 1(b).  The situation is worse at smaller $\langle n \rangle$.  In Fig. 2
we
show the result for
$\langle n \rangle = 1$ in (\ref{9}), for which (\ref{6}) and (\ref{8}) are
again
used.  Notice that there is no connected region of $q$ in which $F_q$ = 1, not
even between $q = 1$ and 2.  Between $q = 9$ and 10 the peak of $|F_q|$ is
greater than $10^4$.  From these results it is therefore not possible to claim
that (\ref{8}) contains no statistical contribution at noninteger $q$.  The
reason for studying $F_q$ at noninteger $q$ is consequently lost.  The $G_q$
moments
\cite{hw,hw2} would be better.

\section{The Solution}\label{III}

To achieve our aim of retaining only the dynamical fluctuation in our
continuation procedure, we demand that (\ref{5}) defines $f(q)$ for all $q$.
Hereafter, we use the notation that when $q$ appears as an argument of a
function, instead of as a subscript, it is to be regarded as a continuous
(complex) variable.  A consequence of that condition is that the normalized
factorial moment function satisfies
  \begin{eqnarray}
F(q) = 1 \label{10}
  \end{eqnarray}
for all $q$ in the case of Poisson distribution.  Equation (\ref{10}) should be
used as a test of the continuation procedure.

The burden of this procedure is to determine $D(t)$.  If a theory specifies
the dynamical distribution $D(t)$ completely, then (\ref{5}) prescribes a
unique continuation of $f_q$ to the complex function $f(q)$ at any $q$.  But
how is that to be checked by experiments where only $P_n$ is
measured?  Thus the procedure must supplement (\ref{5}) with a way of
determining $D(t)$ from $P_n$.  This deconvolution process also cannot be
made unique.  The discrepancies show up as deviations of $F(q)$  from
1.  The region where (\ref{10}) fails significantly can fortunately be
controlled and pushed to large $|q|$.

To deconvolute (\ref{3}) one could consider making the inverse Laplace
transform of the generating function $G(z)$.  However, there are
difficulties connected with the fact that $G(z)$ determined from the
experimental $P_n$ is a polynomial having no singularities in the finite $z$
plane.

Our proposal is to expand $P_n$ in terms of negative binomial distributions
(NBD) $P^{NB}_n(j)$.  One of the attributes of NBD is that it can also be
expressed as a Poisson transform \cite{cfs}, as in (\ref{3}).  They do not
form a complete set of orthogonal functions, so in general they cannot be the
basis functions for the expansion of an arbitrary function.  However, we do
not have an arbitrary function.  The experimental $P_n$ (ignoring errors
for the moment) is a set of $N + 1$ numbers for $n = 0, 1, \cdots, N$.  Thus
the expansion
\begin{eqnarray}
P_n =   \sum ^{N}_{j=0} \,  a_j \, P^{NB}_n ( j)
\label{11}
  \end{eqnarray}
is well defined with $N + 1$ coefficients $a_j$, provided we specify
$P^{NB}_n(j)$ appropriately.  One could consider other distributions instead
of NBD, but for factorial moments that we shall eventually calculate NBD is
most convenient.

Now $P^{NB}_n(j)$ is defined by \cite{cfs}
  \begin{eqnarray}
P^{NB}_n \left( k_j, x_j\right) = {\Gamma (n + k_j) \over \Gamma (n + 1)\Gamma
(k_j)} \left({ k_j \over  k_j + x_j} \right)^{k_j}\,\left({ x_j \over  k_j +
x_j}
\right)^{n}\, ,
\label{12}
  \end{eqnarray}
where $x_j$ and $k_j$ specify the mean and inverse width of $P^{NB}_n(j)$.
How they depend on $j$ will be discussed in the next section.  We remark
that
$x_j$ is equal to
   \begin{eqnarray}
 \bar{n}{(j)} =  \sum^{\infty}_{n=0} n \, P^{NB}_n ( j)
\label{13}
   \end{eqnarray}
only if the sum extends to $\infty$.  Since our method is designed with
phenomenological analysis in mind, where $P_n$ is given only for $n = 0,
\cdots, N,$ all sums over $n$ will be from $0$ to $N$, whether the summand
involves $P_n$ or $P^{NB}_n$.  Consequently, $x_j$ is not {\it exactly}
$\bar{n}(j)$.  Although the discrepancy is small for the $x_j$ and $k_j$ to be
chosen, accuracy will be important, as we shall see.   Extending the sum in
(\ref{11}) to a larger upper limit $N^{\prime}$ with $P_n = 0$ for $N + 1 \leq
n \leq N^{\prime}$ would cause $a_j$ to be very large and highly sensitive to
the accuracy of the calculation; it is a procedure that should  be avoided.
Hereafter $N$ will always be the maximum value of $n$ for which $P_n$ is
measured to be nonzero,  and $x_j$ and $k_j$ should only be regarded as
real parameters of $P^{NB}_n ( j)$ that will be varied in the expansion in
(11).  It should be recognized that because  $P^{NB}_n ( j)$ are all positive
(unlike a harmonic function) and small on the wings, $a_j$ will have
alternating signs and can have large absolute values if $P_n$ is small for a
range of large $n$ values.  Thus accuracy in the ensuing calculations will be
essential.  With
$P^{NB}_n ( j)$ specified, (\ref{11}) is a set of $N + 1$ simultaneous
algebraic equations that can be solved for $a_j$ in terms of the
experimental $P_n$.

Define $D^{NB}(t)$ by the negative-binomial versions of (\ref{3}), i.e.,
  \begin{eqnarray}
P^{NB}_n ( j)  = \int^{\infty}_0 dt \, {t^n \over n!} \, \, e^{-t} \,
D^{NB}(t,j)
\,  . \label{14}
  \end{eqnarray}
Then it is known that \cite{cfs}
  \begin{eqnarray}
D^{NB}(t,j) =  \left({ k_j \over   x_j} \right)^{k_j}  {t ^{k_j - 1} \over
\Gamma
(k_j)} \, e^{- k_j t / x_j }
\,  .
\label{15}
  \end{eqnarray}
The substitution of (\ref{3}) and (\ref{14}) to the two sides of (\ref{11})
results in
  \begin{eqnarray}
D(t) =  \sum^{N}_{j=0}\,  a_j  \, D^{NB}(t,j)\, .
\label{16}
  \end{eqnarray}
Thus we have inverted (\ref{3}) and extracted the dynamical distribution
$D(t)$ from the experimental data on $P_n$.  In principle, this $D(t)$ can be
compared directly with the theoretical distribution. However, the more
familiar arena for comparison involves the factorial moments, which are
more closely related to the data.

An interesting side remark that can be made here is that the feasibility of
determining $D(t)$ from the data makes possible an experimental look at such
theoretical quantity as the Landau free energy if the data on hadronic
multiplicity distribution correspond to quark-hadron phase transition
\cite{hn}, or if the data are on photon distribution at the threshold of lasing
in quantum optics \cite{my}.  That is because in such problems $D(t)$ in
(\ref{3}) is
$e^{-F[t]}$, where $F[t]$ is the free energy of the system.

Returning to the problem on the factorial moments, we substitute (\ref{16})
into our basic equation, (\ref{5}), for continuation to complex $q$, and get
  \begin{eqnarray}
f(q) =  \sum^{N}_{j=0}\,  a_j  \, f^{NB}(q,j)
 \label{17}
     \end{eqnarray}
where
  \begin{eqnarray}
 f^{NB}(q,j)  = \left({ x_j \over k_j}\right)^q \,\, {\Gamma (q + k_j) \over
\Gamma
(k_j)}
\, .
 \label{18}
     \end{eqnarray}
This expression is obtained by integration over $t$ in (\ref{5}) and is valid
only for
   \begin{eqnarray}
\mbox{Re} \, q > - k_j \, .
 \label{19}
     \end{eqnarray}
Thus the domain of $q$ that can be continued into before encountering the
first singularity is governed by the smallest value of $k_j$.

It should be noted that whereas $f_q$ as defined in (\ref{4}) is obtained
directly from the input $P_n$, $f(q)$ is determined from (\ref{17}) after the
inversion is done and the continuation procedure followed.  Even at positive
integer values of
$q$, $f(q)$ is not exactly equal $f_q$ because of finite accuracy in the
computation.  Following (\ref{6}), we define the continued normalized
factorial moments by
   \begin{eqnarray}
F(q) =  f(q) / f(1)^q    \, ,
 \label{20}
     \end{eqnarray}
instead of $f(q)/\langle n \rangle^q$. From (\ref{17}) and (\ref{18}) we then
have
   \begin{eqnarray}
F(q) =  \sum^{N}_{j=0}\,  a_j  \left[{ x_j \over f(1)}\right]^q\, F^{NB}(q,j)
 \label{21}
     \end{eqnarray}
where
			\begin{eqnarray}
F^{NB}(q,j) =  {\Gamma (q + k_j) \over \Gamma
(k_j){k_j} ^q} \, .
 \label{22}
     \end{eqnarray}
Equation (\ref{21}) is our result, which is explicit once the values of $a_j$
are determined.
\section{The Range of $x_j$ and $k_j$}\label{IV}
To complete the description of our continuation procedure, it is necessary to
specify $x_j$ and $k_j$, which represent the average and inverse width of
$P^{NB}_n ( j)$.  They should be chosen to be not too far from the values of
$x$ and $k$ of the input $P_n$, where
\begin{eqnarray}
x \equiv \left< n\right> = \sum^{N}_{n=0}\,n\, P_n \, ,
 \label{23}
     \end{eqnarray}
			\begin{eqnarray}
k \equiv \left( F_2 - 1\right)^{-1} \, , \quad \quad F_2 = \left< n (n -
1)\right>/x^2 \, .
 \label{24}
     \end{eqnarray}
We therefore define first
			\begin{eqnarray}
\Delta _j = \Delta \left( - {1 \over 2} + {j \over N}\right) \, ,
 \label{25}
     \end{eqnarray}
which ranges from $-\Delta/2$ to $+\Delta/2$ in equal steps, as $j$ varies
from $0$ to $N$.  For $x_j$ and $k_j$ to vary from the lower to higher sides
of
$x$ and $k$, we set
		\begin{eqnarray}
x_j = x \left(1 + \Delta _j \right)  \, ,\label{26}
     \end{eqnarray}
		\begin{eqnarray}
k_j = k \left(1 + \Delta _j \right)  \, .
\label{27}
     \end{eqnarray}
Thus $P^{NB}_n ( j)$ with lower $x_j$ has wider width (lower $k_j$); while
that with higher $x_j$ has narrower width.  If  $\Delta$ is not too large, all
components $P^{NB}_n ( j)$ will have comparable magnitudes at large $n$,
which should be small where $P_n$ is small.  Otherwise,  $a_j$  would be
hypersensitive to the accuracy of the computation.  With $\Delta$ being
free to choose, the procedure is obviously not unique.  As we have
remarked earlier, no unique continuation should be expected from a finite
set of numbers, $P_n$.  However, there are guidelines for an optimal choice of
$\Delta$.

We have mentioned in connection with  (\ref{19}) that the domain of
continuation of $q$ is limited by the smallest value of $k_j$, which is $k(1
- \Delta/2)$.  Thus to increase that domain we would want to have a small
value for $\Delta$.  However, with a small range of $x_j$ and $k_j$ it will be
necessary to have highly accurate expansion coefficients $a_j$ in (\ref{11})
to well represent $P_n$.  So a larger value of $\Delta$ is preferred.  This
point needs to be demonstrated quantitatively.  A way to do this is to
calculate
$f(1)$ and examine its dependence on $\Delta$.  To that end we consider
specific examples in the following.

Consider a sample $P_n$ given by
		\begin{eqnarray}
{\cal P}^{(1)}_n = \left( n + 0.3\right) ^3 \, e^{- 0.5 n }/ Z
,\qquad \qquad  n= 0, \cdots, N
 \label{28}
     \end{eqnarray}
where $N = 30$ and $Z$ is the normalization factor so that $\sum^N_{n =
0}{\cal P}^{(1)}_n = 1$.  For this ${\cal P}^{(1)}_n $ we have $x = \langle n
\rangle = 7.6966$ and $k = 7.2193$.  For $\Delta$ chosen to be in the range
$0.1 \leq \Delta \leq 0.6$, we follow the procedure described in Sec. \ref{III}
and calculate $f(1)$.  The result is shown in Fig.\ 3(a) as a function of
$\Delta$.  Evidently, there are fluctuations at small values of $\Delta$, but
quite stable for $\Delta \geq 0.25$.  However, when the vertical scale is
expanded as shown in Fig. 3(b), we see that small fluctuations at a level
of $<$ 0.3\% are still present until  $\Delta \geq 0.4$, where $f(1) = 7.7016$.
A discrepancy between $f(1)$ and $x$ is anticipated because of the finiteness
of $N$  but at 0.06\% level it is unimportant.  What is important is that
$\Delta$ should not be too small.  This example demonstrates the limitation
of the method due to (\ref{19}), if one attempts to continue $q$ to large
negative values.  From Fig.\ 3 an appropriate value for $\Delta$ can be set at
0.5, for which min$\{k_j\} = 0.75 \, k = 5.4145$.  Thus $F(q)$ can be
continued to $q
\simeq - 5.4$ before encountering divergence.  In practice this range of
negative $q$ is quite enough to exhibit the low $n$ behavior of $P_n$.
Increasing $\Delta$ would improve the stability of the solution, but decrease
the range of continued $q$.  The choice of $\Delta = 0.5$ seems like a good
compromise between the two opposite preferences.

Consider next another example where $\langle n \rangle$ is much smaller,
corresponding
to the situation where the phase-space cell size $\delta$ is small.  Assume
			\begin{eqnarray}
{\cal P}^{(2)}_n = \left( n + 1\right) ^{0.5} \, e^{-  n }/ Z ,
\qquad  \qquad  N = 20 \, ,
 \label{29}
     \end{eqnarray}
for which $x = 0.8352$ and $k = 1.395$.  Fig.\ 4 shows the result of $f(1)$
vs $\Delta$, which is rather free of fluctuation for all $\Delta$ except near
0.1.  The value of $f(1) = 0.8352$ is equal to $x$ to 5 significant figures.
Choosing $\Delta = 0.5$ gives min$\{k_j\} = 1.046$ which does not allow $q$
to go much beyond $-1$. Decreasing $\Delta$ would not improve the situation
due to the limitation of small $k$ for ${\cal P}^{(2)}_n$.  Since $\Delta$
should
not be changed in the analysis of data with varying $\delta$, we suggest that
$\Delta$ be fixed at 0.5.

\section{Continued Factorial Moments}\label{V}

The continuation procedure having been completely specified in Secs.
\ref{III} and \ref{IV}, we can now proceed to the study of the normalized
factorial moments $F(q)$.   We continue to use the three distributions
${\cal P}^{(i)}_n$, $i = 0, 1, 2,$ as our sample inputs for $P_n$.  Unless
otherwise stated, $\Delta = 0.5$  is used.

For the Poisson distribution ${\cal P}^{(0)}_n$, let us consider the same two
cases:  (a) $\langle n \rangle = 6$ and (b) $\langle n \rangle = 1$, already
examined in Sec. \ref{II},
where the simple continuation scheme $n! \rightarrow \Gamma (n + 1)$
was used .  Now, we assume $N = 30$ for (a) and $N = 10$ for (b) as the
upper limits of $n$ in the experimental $P_n$.  If $N = \infty$, ${\cal
P}^{(0)}_n$ would give $F_2 = 1$, so according to (\ref{24}) $k$ would be
infinite. For the finite $N$ chosen for the two cases, $F_2$ are still very
close to 1, so $k$ would be extremely large.  For our calculation it is
sufficient to set $k = 10^4$.  Using our procedure of continuation the results
are shown in Fig.\ 5(a) and (b), for the two cases (a) and (b), respectively.
Note the high resolution of the vertical scale.
For $\langle n \rangle = 6$ in case (a) $F(q)$ is essentially 1 for $-10 < q <
30$.
In case (b) where $\langle n \rangle = 1$, $F(q)$ is
almost 1 for $-10 < q < 20$ except near the edges of that range.  This case
 is more difficult to continue accurately because there are fewer
values of nonvanishing
${\cal P}^{(0)}_n$.  Theoretically, ${\cal P}^{(0)}_n$ is nonzero for any
finite $n$,
but we cut off at $N = 10$ to  simulate a realistic situation where $\langle n
\rangle$ is only 1.  With only 11 values of $a_j$ the continuation to large
$|q|$ cannot be expected to have extremely high accuracy.  That is why
$F(q)$ deviates from 1 near the two ends in Fig.\ 5(b). Nevertheless, the
deviation is only of order 0.1\%.  Upon comparing Fig.\ 5 to Figs.\ 1 and 2,
the advantage of this method over the simple scheme of Sec. \ref{II} is
self-evident.

For ${\cal P}^{(1)}_n$ given in (\ref{28}) with $N = 30$ the result of
our calculation for $F(q)$ is shown in Fig.\ 6 (a) and (b).  The dependence on
$q$ is evidently very smooth.  It grows rapidly at negative $q$, even though
the first singularity is located at $q < -5$.  We have also calculated $F(q)$
for
$\Delta = 0.3$, the result of which is plotted in dashed lines, lying very
close
to the solid lines for $\Delta = 0.5$.  For $q > 0$ the two cases are
indistinguishable.  For $q < -2$ the difference is actually not negligible in
absolute value but because of the rapid rise of $F(q)$ it is not significant in
terms of percentage discrepancy.  In Fig.\ 6(b) we see that for the range of
$q$ shown the difference between the two cases is totally insignificant.  This
is the most important range for the continuous $q$ problem, and we have
found a reliable continuation of $F(q)$.

It should be pointed out that $F(0)=1$ is not accidental.  From (21) and (22)
we see that $F(0)=\Sigma_j a_j$, which is 1 by virtue of the normalization of
$P_n$ and $P_n^{NB}(j)$ in (11).  Of course, $f_0$ is also 1 if the
continuation
scheme of (8) is followed.

Finally, let us come to the third example where ${\cal P}^{(2)}_n$ is as given
in (\ref{29}) with $N = 20$.  Now, the values of $x$ and $k$ are small.  The
calculated result for $F(q)$ is shown in Fig.\ 7 (a) and (b).  There is a fast
rise
at large $q$ because of the smaller $x$ (compared to the case above).  The
continuation to negative $q$ encounters irregularity due to the small value
of $k$.  For $q > -0.5$, there is essentially no difference between the use of
$\Delta = 0.5$ (solid line) and $\Delta = 0.3$ (dashed line).  Only the solid
line is plotted in Fig.\ 7(a); both are plotted in Fig.\ 7(b).  The difference
between the two $\Delta$ cases becomes noticable and quantitatively
significant only for $q < -0.5$, a region very close to the singularities.  The
reliability of our continuation should therefore be restricted to the domain
to the right of the sudden downturn of $F(q)$ around $q = -0.5$.  In that
domain our result is smooth and insensitive to $\Delta$.

\section{Multifractal Analysis}\label{VI}

With $F(q)$ continuable to noninteger $q$, it is now possible to consider
multifractal analysis, assuming that $F(q)$ has a power-law dependence on
the resolution scale $\delta$ for a range of $q$ covering both positive and
negative values.  Such an analysis was suggested previously using $G$
moments, which are defined for all $q$ by \cite{hw}
     \begin{eqnarray}
G(q) = \sum _i \left({n_i  \over n_t }  \right)^q
 \label{30}
     \end{eqnarray}
where $n_i$ is the multiplicity in bin $i$, $n_t = \sum _i n_i$, and the sum in
$i$ is over all nonempty bins.  $G(q)$ shows scaling behavior
		\begin{eqnarray}
G(q) \sim \delta^{\tau(q)}
 \label{31}
     \end{eqnarray}
for $q > 1$ in both experimental data and model simulation, when
$\theta(n_i - q)$ is included in the summand in (\ref{30}) \cite{hw2}.
However, for $q < 1$ (\ref{31}) is not valid for any extended range of
$\delta$, so multifractal analysis cannot be made for that range of $q$.  The
problem is rooted in the empty-bin effect and the fact that $G(q)$ contains
statistical fluctuations.  Since $F(q)$ is now defined for noninteger $q$, its
scaling behavior
		\begin{eqnarray}
F(q) \propto \delta^{-\varphi(q)} \, ,
 \label{32}
     \end{eqnarray}
especially for negative $q$, should be checked for a variety of existing data.
If (\ref{32}) is valid for a range of $q$ around 1, multifractal analysis can
then proceed without the necessity of subtracting out the statistical
component, as was done for $G(q)$.

We can relate $G(q)$ and $F(q)$, if we assume that (\ref{30}) is defined for
the dynamical distribution only without the statistical fluctuation.  The sum
over all bins in (\ref{30}) can then be related to averaging over the
dynamical distribution in (\ref{5}).  Using (\ref{6}), (\ref{31}) and
(\ref{32}), we get
\begin{eqnarray}
\tau(q) = q - 1 - \varphi(q) \, ,
 \label{33}
     \end{eqnarray}
where the $-1$ comes from the fact that $\sum_i$ by itself gives the total
number of bins, which varies as $\delta ^{-1}$.  Multifractal spectrum is then
obtained by the Legendre transform \cite{hw,fed}
	\begin{eqnarray}
f(\alpha) = q \alpha - \tau(q)
 \label{34}
     \end{eqnarray}
with
\begin{eqnarray}
\alpha = d \tau(q)/dq \quad .
 \label{35}
     \end{eqnarray}
Thus the verification of the power law (\ref{32}) and the capability of
calculating $\alpha$ by differentiation with respect to $q$, which we now
have, make possible the presentation of the scaling properties of dynamical
fluctuations in terms of the multifractal spectrum $f(\alpha)$.

It is not our purpose here to examine specific dynamical models or
experimental data and to extract their multiplicity fluctuation behaviors.
However, we can demonstrate the nature of  $f(\alpha)$ if we assume that
the scaling behavior (\ref{32}) is true for a set of $P_n(\delta)$ for a range
of $\delta$.  Let us further assume that among those $P_n(\delta)$, a
specific one at some $\delta_0$ is exactly ${\cal P}_n^{(1)}$ given in
(\ref{28}).  Then we have
\begin{eqnarray}
\varphi(q) = c \, {\rm ln} \,  F(q) + c_1(q) \quad ,
 \label{36}
     \end{eqnarray}
where $c = (- {\rm ln} \, \delta_0 )^{-1}$ and $c_1(q)$ is some function of
$q$ independent of $\delta_0$ arising from the proportionality factor in
(\ref{32}).  Scaling behavior means that $\varphi (q)$ is unchanged, as
${\delta}$ is varied from $\delta_0$. The major part of the $q$ dependence
of $\varphi (q)$ derives from that of $F(q)$, which we know from
Fig.\ 6.  Apart from the unknown constant $c$ and the unknown function
$c_1(q)$ in this example, we can determine $f(\alpha)$ from $F(q)$ by
varying $q$ parametrically.   For illustrative purpose, let us assume that
$c_1(q) = 0$ so that using  (\ref{33})-(\ref{36}) and the result of our
calculated $F(q)$ pertaining to
${\cal P}_n^{(1)}$, we can determine $f(\alpha)$.  In Fig.\ 8 we show the
result for four possible values of $c$.  We stress that in a model or data
analysis
$c$ is not a variable;  $\varphi (q)$ is determined from the log-log plots
when there is scaling.  Fig.\ 8 merely illustrates the possible form of
$f(\alpha)$, if
$\varphi (q)$ happens to coincide with the result of analyzing a particular
$P_n = {\cal P}_n^{(1)}$ with a specific $c$ in (\ref{36}) and with $c_1(q) =
0$.  The dashed line indicates where $f(\alpha) = \alpha$, and is tangent to
each of the
$f(\alpha)$ curves at $q = 1$.  The range of $q$ covered by $f(\alpha)$ in
Fig.\ 8 is, depending on $c$, roughly $-2 \leq q \leq 4$, with the $q =
0$ point always occurring at the peak of the $f(\alpha)$ curve.  The
multifractal dimension $D_q$ is \cite{hw,hen}
\begin{eqnarray}
D_q = \tau(q) / (q - 1) \quad ,
 \label{37}
     \end{eqnarray}
which is related to $\alpha$ by
		\begin{eqnarray}
D_0 = f(\alpha _0) \quad , \quad \quad  D_1 = \alpha _1 \quad ,
 \label{38}
     \end{eqnarray}
where $\alpha_q$ is the value of $\alpha$ at $q$.  Thus $\alpha_0$ is
where $f(\alpha)$ is maximum, and $\alpha _1$ is where $f(\alpha _1) =
\alpha _1$.  The multifractal spectrum $f(\alpha)$ is the most elegant way
of displaying the scaling properties of dynamical fluctuation.  Reduced to the
bare minimum, two parameters can be used to characterize $f(\alpha)$, viz.
$\alpha_0$ and $\alpha _1$, the location of the peak and a measure of the
width, respectively.

In many multiparticle production processes the scaling behavior (\ref{32}) is
not valid over an extended range of $\delta$.  The multifractal analysis
described above cannot be applied then.  However, it has been found
phenomenologically \cite{oc,cha,ph} as well as theoretically \cite{hn,hw4}, not
only in hadronic and nuclear collisions, but also in quantum optics \cite{my},
that $F_q$ satisfies a different scaling law
		\begin{eqnarray}
F_q \propto F^{\beta_q}_2 \quad .
 \label{39}
     \end{eqnarray}
Let us assume that this behavior can be established for continuous $q$ so
that the function $\beta(q)$ can be determined for a range of $q$ values
both positive and negative.  Then it is possible to define formally another
spectrum, call it $g(\alpha)$, in exact analogy to  (\ref{33})-(\ref{35}), but
without the geometrical implication of multifractality.  Thus we define
		\begin{eqnarray}
\sigma(q) = q - 1 -  \beta(q)
\label{40}
     \end{eqnarray}
		\begin{eqnarray}
g(\alpha) = q \alpha  -  \sigma(q)
 \label{41}
     \end{eqnarray}
		\begin{eqnarray}
\alpha = d \sigma(q) / dq \quad .
 \label{42}
     \end{eqnarray}
The only input into this scheme of description is $\beta(q)$, which is
\begin{eqnarray}
\beta(q) = {\rm ln}\,  F(q) / {\rm ln}\,  F(2) + b(q) \, ,
 \label{43}
     \end{eqnarray}
where $b(q)$ is the log of the proportionality factor in
(\ref{39}).  Mathematically,
$\sigma (q)$ and
$g(\alpha)$ correspond to
$\tau (q)$ and
$f(\alpha)$ if we set $c = 1/ {\rm ln} \, F(2)$, but physically (\ref{32})
need not be true, rendering $\varphi(q)$ meaningless, while (\ref{39}) can
well be true (no exception having been found so far).  Ref. \cite{hn} gives an
explicit example of (\ref{39}) being valid for a problem that does not have
(\ref{32}).

In the example where ${\cal P}_n^{(1)}$ is considered, let us assume that it
belongs to the type of physical problems for which (\ref{39}) is valid.  Then
the function $F(q)$ obtained is sufficient to determine $g(\alpha)$,
assuming $b(q) = 0$.  The result is shown in Fig.\ 9(a).  Since $F(2) = 1.14$,
$g(\alpha)$ corresponds to $f(\alpha)$ in Fig.\ 8 with c = 7.64.  In Fig.\ 9(a)
the peak occurs at $\alpha _0 = 1.45$ while the tangent point is at $\alpha
_1 = 0.52$.  If, on the other hand, ${\cal P}_n^{(2)}$ is used as an example
that has a scaling behavior (\ref{39}), totally unrelated to ${\cal
P}_n^{(1)}$,
then the corresponding spectrum $g(\alpha)$ is as shown in Fig.\ 9(b).  Note
that in this case the maximum $\alpha$ is 2.15 corresponding to $q = -
0.24$.  Our continuation to lower value of $q$ has led to a sudden downturn
of $F(q)$ around $q = -0.4$, exhibited in Fig.\ 7(b).  That causes a drastic
change in the derivative in (\ref{42}) at around $q = -0.24$ which in turn
gives rise to an irregular behavior in $g(\alpha)$ at $\alpha =
2.15$.  Thus the nature of ${\cal P}_n^{(2)}$ prevents the use of $F(q)$ for
$q < -0.24$, thereby setting an upper bound to how far to the right the
spectrum $g(\alpha)$ can be developed.  In that figure $\alpha _0 = 1.76$
and  $\alpha _1 = 0.41$, quite different from the corresponding values in
Fig.\ 9(a).  These two figures are sufficient to indicate that ${\cal
P}_n^{(1)}$
and
${\cal P}_n^{(2)}$ cannot belong to the same class of $F$-scaling factorial
moments.

In summary, $g(\alpha)$ is a representation of the
characteristics of the $F$-scaling behavior, (\ref{39}), of $F(q)$, and may be
a generally useful description of all multiparticle production processes.

\section{Conclusion}\label{VII}

We have presented a way to determine $F(q)$ for continuous $q$ such that
it is 1 for all $q$ if the input distribution is Poissonian, i.e. the
statistical
fluctuation is filtered out.  The range of $q$ into which $F(q)$ can be
continued depends on the nature of $P_n$.  Generally speaking, the
low $n$ part of $P_n$ is characterized by the negative $q$ region of $F(q)$.
Thus the study of the scaling behavior of multiplicity fluctuations can now
be extended to dips, gaps and voids.  All existing data that have been put
to intermittency analysis at positive integer $q$ should be reanalyzed for
continuous $q$.  Similarly, such reanalysis should be done for all models and
MC codes.  Thus the confrontation between theory and experiment can now
be extended to a significant portion of the real line of $q$, as compared to
the previous situation where it has been done for only a few isolated points
at the integer values.  For comparison, the study of Bose-Einstein correlation
is focused on only one point: $q = 2$.

If the dynamics of particle production is self-similar so that $F(q)$ exhibits
a power-law dependence on the resolution scale $\delta$, then the
intermittency index $\varphi(q)$ can be determined as a continuous
function of $q$.  It then follows that the multifractal spectrum $f(\alpha)$
can be derived without any ambiguity or need for correction to eliminate
statistical contamination.  On the other hand, if there is no power-law
dependence on $\delta$, but there is $F$-scaling, i.e., $F_q \propto
F^{\beta_q}_2$, which is more commonly observed, then the knowledge of
$\beta(q)$ is sufficient to determine the spectrum $g(\alpha)$ that gives
an excellent description of the self-similar behavior of the dynamics of
fluctuations.

For the purpose of providing a convenient outline of the procedure to
determine $F(q)$, $f(\alpha)$ and $g(\alpha)$, we summarize here the steps
needed to do the analysis:

\begin{enumerate}

\item  Starting with the input $P_n$, $n = 0$, $\cdots$, $N$, determine $x$
and $k$ from (\ref{23}) and (\ref{24}), and then $x_j$ and $k_j$ from
(\ref{25})-(\ref{27}) with $\Delta = 0.5$.

\item Using (\ref{12}), set up $N + 1$ linear algebraic equations, (\ref{11}),
with $P^{NB}_n ( j)$ as the matrix and $P_n$ as the input vector. Solve for
$a_j$, $j = 0$, $\cdots$, $N$.  [The use of MATHEMATICA has been found to
be convenient.]

\item Use (\ref{21}) to calculate $F(q)$ for a range of real $q$.

\item  If the input $P_n (\delta)$ is known for a range of $\delta$,
determine $F(q)$ as a function of $\delta$ and examine the validities of
(\ref{32}) and (\ref{39}) over that range of $\delta$.

\item If (\ref{32}) is valid for a subrange of $\delta$, then from
$\varphi(q)$ and (\ref{33})-(\ref{35}) determine $f(\alpha)$.

\item If (\ref{32}) is invalid but (\ref{39}) is valid, then from $\beta(q)$
and (\ref{40})-(\ref{42}) determine $g(\alpha)$.

\item  Compare theory and experiment at any of the three levels:  $F(q)$,
$f(\alpha)$ or $g(\alpha)$.

\end{enumerate}
\vskip 1.5cm
\centerline {ACKNOWLEDGMENTS}
\vskip .5cm
I am grateful to  I.\ M.\ Dremin and J.-L.\ Meunier for helpful
discussions.   This work was supported, in part, by the U.\ S.\ Department of
Energy under Grant No.\ DE-FG06-91ER40637.

\newpage

\vskip .5cm
\begin{center}
\section*{Figure Captions}
\end{center}

\begin{enumerate}

\item[Fig.\ 1] Normalized factorial moments $F_q$ in the simple continuation
procedure using (\ref{8}) with Poisson distribution (\ref{9}) as input and
with $\langle n \rangle = 6$.  (a) $-10 \leq q \leq 25$; (b) $25 \leq q \leq
30$.

\item[Fig.\ 2]Same as Fig.\ 1 but for $\langle n \rangle = 1$.

\item[Fig.\ 3]Dependence of factorial moment $f(q = 1)$ on $\Delta$ when the
input distribution is ${\cal P}_n^{(1)}$.  The resolution in (b) is higher than
that in (a).

\item[Fig.\ 4]Dependence of  $f(q = 1)$ on $\Delta$ when the
input distribution is ${\cal P}_n^{(2)}$.

\item[Fig.\ 5]Normalized factorial moments $F(q)$ with high vertical
resolution for Poisson distribution with (a) $\langle n \rangle = 6$, and (b)
$\langle n \rangle = 1$.

\item[Fig.\ 6]$F(q)$ with ${\cal P}_n^{(1)}$ as input distribution.  Solid line
is
for $\Delta = 0.5$ dashed line for  $\Delta = 0.3$. (a) $-3 \leq q \leq 10$;
(b) $-1.5 \leq q \leq 2.5$.

\item[Fig.\ 7]$F(q)$ with ${\cal P}_n^{(2)}$ as input distribution.  Solid line
is
for $\Delta = 0.5$ dashed line for  $\Delta = 0.3$. (a) $-1 \leq q \leq 6$;
(b) $-1 \leq q \leq 2$.

\item[Fig.\ 8]Multifractal spectra $f(\alpha)$ for four values of $c$ (see
text),
if the input ${\cal P}_n^{(1)}$ belongs to a class of scaling distributions.
Dashed line is for $f(\alpha) = \alpha$.

\item[Fig.\ 9]Spectrum function $g(\alpha)$ if (a) ${\cal P}_n^{(1)}$ and (b)
${\cal P}_n^{(2)}$ belong to separate classes of self-similar distributions
satisfying $F$-scaling (3.9).

\end{enumerate}

\end{document}